\documentclass[12pt]{iopart}
\usepackage{epsf}
\begin{document}
\title[Gravitational waves from PSR J1939+2134]{Upper limits on
the strength of periodic gravitational waves from PSR J1939+2134}
%
%
%
\renewcommand{\footnote}[1]{}
\newcommand*{\AG}{$^{1}$}
\newcommand*{\AH}{$^{2}$}
\newcommand*{\AN}{$^{3}$}
\newcommand*{\CH}{$^{4}$}
\newcommand*{\DO}{$^{5}$}
\newcommand*{\CA}{$^{6}$}
\newcommand*{\CU}{$^{7}$}
\newcommand*{\CL}{$^{8}$}
\newcommand*{\CO}{$^{9}$}
\newcommand*{\FN}{$^{10}$}
\newcommand*{\HC}{$^{11}$}
\newcommand*{\IU}{$^{12}$}
\newcommand*{\CT}{$^{13}$}
\newcommand*{\LM}{$^{14}$}
\newcommand*{\LO}{$^{15}$}
\newcommand*{\LV}{$^{16}$}
\newcommand*{\LU}{$^{17}$}
\newcommand*{\LE}{$^{18}$}
\newcommand*{\LL}{$^{19}$}
\newcommand*{\MP}{$^{20}$}
\newcommand*{\ND}{$^{21}$}
\newcommand*{\NA}{$^{22}$}
\newcommand*{\NO}{$^{23}$}
\newcommand*{\SC}{$^{24}$}
\newcommand*{\SE}{$^{25}$}
\newcommand*{\SA}{$^{26}$}
\newcommand*{\SR}{$^{27}$}
\newcommand*{\PU}{$^{28}$}
\newcommand*{\TC}{$^{29}$}
\newcommand*{\TR}{$^{30}$}
\newcommand*{\HU}{$^{31}$}
\newcommand*{\BB}{$^{32}$}
\newcommand*{\BR}{$^{33}$}
\newcommand*{\FA}{$^{34}$}
\newcommand*{\GU}{$^{35}$}
\newcommand*{\MU}{$^{36}$}
\newcommand*{\OU}{$^{37}$}
\newcommand*{\RO}{$^{38}$}
\newcommand*{\UW}{$^{39}$}
\newcommand*{\WU}{$^{40}$}
\author{
B.~Allen\UW and G.~Woan\GU, for the LIGO Scientific Collaboration:
\scriptsize
B.~Abbott\CT,
R.~Abbott\CT,
R.~Adhikari\LM,
B.~Allen\UW,
R.~Amin\FA,
S.~B.~Anderson\CT,
W.~G.~Anderson\TC,
M.~Araya\CT
% ...
}
\address{\AG Albert Einstein Institut f\"ur Gravitationsphysik, D-14476 Golm, Germany}
\address{\AH Albert Einstein Institut f\"ur Gravitationsphysik, D-30157 Hannover, Germany}
\address{\AN Australian National University, Canberra, 0200, Australia}
\address{\CH California Institute of Technology, Pasadena, CA  91125, USA}
\address{\DO California State University Dominguez Hills, Carson, CA  90747, USA}
\address{\CA Caltech-CaRT, Pasadena, CA  91125, USA}
\address{\CU Cardiff University, Cardiff, CF2 3YB, United Kingdom}
\address{\CL Carleton College, Northfield, MN  55057, USA}
\address{\CO Cornell University, Ithaca, NY  14853, USA}
\address{\FN Fermi National Accelerator Laboratory, Batavia, IL  60510, USA}
\address{\HC Hobart and William Smith Colleges, Geneva, NY  14456, USA}
\address{\IU Inter-University Centre for Astronomy  and Astrophysics, Pune - 411007, India}
\address{\CT LIGO - California Institute of Technology, Pasadena, CA  91125, USA}
\address{\LM LIGO - Massachusetts Institute of Technology, Cambridge, MA 02139, USA}
\address{\LO LIGO Hanford Observatory, Richland, WA  99352, USA}
\address{\LV LIGO Livingston Observatory, Livingston, LA  70754, USA}
\address{\LU Louisiana State University, Baton Rouge, LA  70803, USA}
\address{\LE Louisiana Tech University, Ruston, LA  71272, USA}
\address{\LL Loyola University, New Orleans, LA 70118, USA}
\address{\MP Max Planck Institut f\"ur Quantenoptik, D-85748, Garching, Germany}
\address{\ND NASA/Goddard Space Flight Center, Greenbelt, MD  20771, USA}
\address{\NA National Astronomical Observatory of Japan, Tokyo  181-8588, Japan}
\address{\NO Northwestern University, Evanston, IL  60208, USA}
\address{\SC Salish Kootenai College, Pablo, MT  59855, USA}
\address{\SE Southeastern Louisiana University, Hammond, LA  70402, USA}
\address{\SA Stanford University, Stanford, CA  94305, USA}
\address{\SR Syracuse University, Syracuse, NY  13244, USA}
\address{\PU The Pennsylvania State University, University Park, PA  16802, USA}
\address{\TC The University of Texas at Brownsville and Texas Southmost College, Brownsville, TX  78520, USA}
\address{\TR Trinity University, San Antonio, TX  78212, USA}
\address{\HU Universit{\"a}t Hannover, D-30167 Hannover, Germany}
\address{\BB Universitat de les Illes Balears, E-07071 Palma de Mallorca, Spain}
\address{\BR University of Birmingham, Birmingham, B15 2TT, United Kingdom}
\address{\FA University of Florida, Gainsville, FL  32611, USA}
\address{\GU University of Glasgow, Glasgow, G12 8QQ, United Kingdom}
\address{\MU University of Michigan, Ann Arbor, MI  48109, USA}
\address{\OU University of Oregon, Eugene, OR  97403, USA}
\address{\RO University of Rochester, Rochester, NY  14627, USA}
\address{\UW University of Wisconsin-Milwaukee, Milwaukee, WI  53201, USA}
\address{\WU Washington State University, Pullman, WA 99164, USA}
\begin{abstract}
The first science run of the LIGO and GEO gravitational wave
detectors presented the opportunity to test methods of searching
for gravitational waves from known pulsars.  Here we present new
direct upper limits on the strength of waves from the pulsar PSR
J1939+2134 using two independent analysis methods, one in the
frequency domain using frequentist statistics and one in the time
domain using Bayesian inference. Both methods show that the strain
amplitude at Earth from this pulsar is less than a few times
$10^{-22}$.
\end{abstract}


\submitto{\CQG}


\section{Introduction}
The LIGO and GEO gravitational wave detectors performed their
first science run, denoted S1, from 23 August to 9 September 2002
(Abbott et al., 2003a, for the LIGO Scientific Collaboration).
Although these instruments were still to reach their design
sensitivities, their performances  during this period were
sufficiently good to justify a serious test of our search
algorithms on real interferometer data.  Code to search for
continuous gravitational waves from rapidly rotating neutron stars
has been under development within the LIGO Scientific
Collaboration (LSC) since the mid-to-late 1990s, and for S1 the
LSC Pulsar Upper Limits Group (PULG) used two independent search
methods to set upper limits on signals from the millisecond pulsar
PSR J1939+2134.  This object has the shortest spin period of any
known pulsar and falls in a relatively quiet part of the detector
bands. Its short period also enables us to place a relatively
tight upper limit on the equatorial ellipticity of the neutron
star.  Here we present a brief outline of the results from this
exercise.  A more detailed presentation of the methods and
discussion of the results can be found in Abbott et al., 2003b
(for the LIGO Scientific Collaboration).

\section{Detector sensitivities}
In setting an upper limit on the gravitational wave amplitude from
a pulsar we are saying something about the pulsar's physical
parameters as constrained by instrumental performance rather than
simply making a statement about the strain sensitivity of our
instrument.  The strength of any continuous gravitational wave
signal from PSR J1939+2134 seen by a detecting interferometer will
depend not only on the pulsar's distance, quadrupole moment and
spin period, but also on the relative orientation between its axis
of spin and the arms of the interferometer.  The diurnal motion of
the pulsar through the antenna pattern will also modulate the
strength of the signal, so there is no simple relationship between
the strain sensitivity curve of an interferometer and the upper
limit that interferometer can place on the parameters of a
particular pulsar.  We can however define a \emph{mean}
performance, averaged over these unknowns, for a particular data
set and detection method.
\begin{figure}
\begin{center}
\epsfxsize=9cm \epsfbox{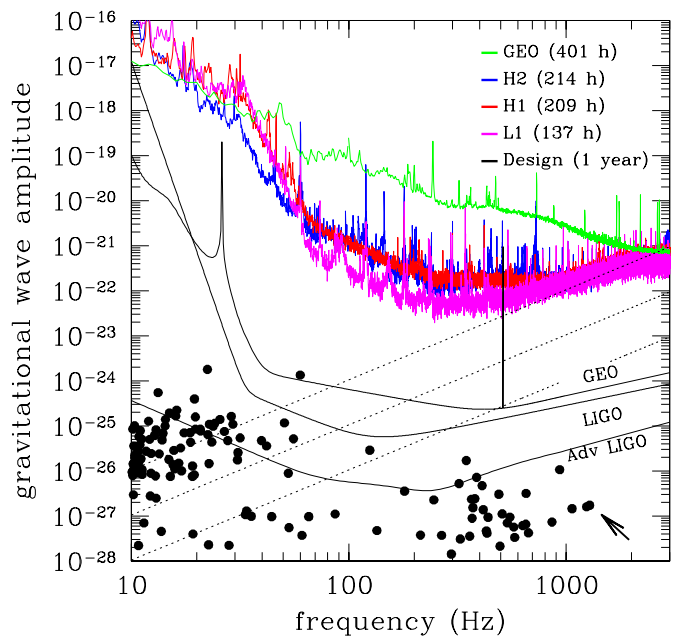}
\end{center}
\vspace*{-0.5cm} \caption{\label{f1}  Mean upper limits, as a
function of signal frequency, for the characteristic amplitude of
gravitational waves from rotating neutron stars as constrained by
the S1 data (upper curves).  The lower three curves represent 1\,y
design sensitivities for GEO, LIGO and one possible design for
Advanced LIGO.  The dotted lines show strain amplitudes
corresponding to neutron stars at 8.5\,kpc with equatorial
ellipticities of $10^{-3}$, $10^{-4}$ and $10^{-5}$.  The dots
show characteristic amplitudes from known pulsars assuming all the
observed loss in rotational kinetic energy is dissipated in
gravitational waves. The arrow points to PSR J1939+2134. }
\end{figure}
Figure~\ref{f1} shows how our interferometers constrain
monochromatic strain amplitudes at Earth as a function of signal
frequency, averaged over the antenna pattern and relative
orientation of the pulsar spin axis to the interferometer.  The
curves are derived from the `frequency domain method' described
below, and represent detection with a 1\% false alarm rate and a
10\% false dismissal rate using the actual observation times and
data from S1.  The lower solid curves show the design
sensitivities of the interferometers for a 1\,y observation
period.  The dotted lines indicate the strain amplitudes
corresponding to hypothetical neutron stars at 8.5\,kpc with
equatorial ellipticities of $10^{-3}$, $10^{-4}$ and $10^{-5}$.
Finally the dots represent the upper limit on strain amplitude for
a selection of pulsars.  For this we have assumed a simple
spindown model in which  all the apparent loss in rotational
kinetic energy of the neutron star goes into the emission of
gravitational waves.  It is clear from this figure that no
detection was expected from the S1 data.  Our primary aims in this
exercise were to test the data reduction and analysis methods in
the presence of realistic interferometer noise and set an upper
limit on the strengths of such waves.

\section{Search methods}
Although the location, period and period derivative of PSR
J1939+2134 are known to high precision the inclination of the
neutron star's spin axis to the line of sight ($\iota$) and the
phase ($\phi$), polarization ($\psi$) and, of course, amplitude
($h_0$) of the gravitational wave are unknown.   All these factors
affect the form of the signal and were searched over in S1 using
two independent methods.

The `frequency domain method' uses Fourier transform techniques to
set a frequentist upper limit on the strain amplitude from the
pulsar.   The input time series is broken down into 60\,s-duration
short Fourier transforms (SFTs) which are high pass filtered at
100\,Hz, Tukey windowed in the time domain and amplitude
calibrated once per minute. These SFTs are then optimally combined
to take account of the Doppler evolution and spindown of the
signal using a Dirichlet kernel method (Jaranowski et al., 1998)
and used to compute a detection statistic, denoted $\cal F$, which
is the likelihood ratio of the data, maximized over $\iota$,
$\phi$ and $\psi$. Monte Carlo injections into the data stream are
then used to calculate the probability distribution function (pdf)
of $\cal F$. The upper limit is defined such that had a signal at
the upper limit level or greater been present in an ensemble of
similar experiments to the one performed, 95\% of the experiments
would yield values of $\cal F$ that exceed the one seen in the
real experiment.  The computing requirement for this Monte Carlo
work is extensive, and the searches ran off-line on the Medusa
cluster at UWM, (296 single 1\,GHz CPU nodes and 58\,TB of disk
space), and the Merlin cluster at AEI Potsdam (180 dual 1.4\,GHz
CPU nodes and 36\,TB of disk space).

The `time domain method' is an algorithm specifically developed to
search for neutron stars with a complex, but known, rotational
phase evolution.  It uses heterodyne methods to reduce and filter
the data to a rate of one complex value every 60\,s, estimating
the noise variance over the same period to take account of
non-stationarity in the data.  The algorithm proceeds to carry out
the search as a standard Bayesian modelling problem, determining a
posterior pdf for the unknown source parameters $h_0$, $\iota$,
$\phi$ and $\psi$.  The prior for $h_0$ is chosen to be uniform
over $[0,\infty]$ and 0 for $h_0<0$, and the three other priors
are chosen to be least informative about the orientation of the
neutron star.  The time domain Bayesian upper limit is set via the
posterior pdf for $h_0$, marginalized over the nuisance parameters
$\iota$, $\phi$ and $\psi$.  We define the 95\% upper credible
limit for $h_0$ as the value at which the cumulative probability
of $h_0$ (from zero) is 0.95.

Note that there is nothing special about the choice of domain
(frequency or time) for the frequentist and Bayesian analyses --
the above choice was influenced by the desire to have two very
different approaches to solving the same problem.

\section{Results}
As expected the S1 data provided no evidence for continuous wave
emission from PSR J1939+2134 at twice its rotation frequency (as
determined by radio observations).  However, we were able to
improve the best upper limit for such signals by about a factor of
100, based on earlier work by Hough et al.\ (1983) and Hereld
(1983).

The results from the four interferometers that took part in S1 are
summarized in Table~\ref{t1}, the quoted uncertainties being
largely due to calibration.  The two analysis methods are very
different, and the meaning of `upper limit' is not the same for
both, however we would hope that the concept of an upper limit is
sufficiently robust that the two interpretations would deliver
numerically similar results, as indeed they do.
\begin{table}
\begin{center}
\begin{tabular}{lcc}
  Instrument & Frequentist UL & Bayesian UL \\ \hline
  GEO    & $(1.9 \pm 0.1)\times 10^{-21}$ & $(2.2 \pm 0.1)\times 10^{-21}$ \\
  LLO    & $(2.7 \pm 0.3)\times 10^{-22}$ & $(1.4 \pm 0.1)\times 10^{-22}$ \\
  LHO-2k & $(5.4 \pm 0.6)\times 10^{-22}$ & $(3.3 \pm 0.3)\times 10^{-22}$ \\
  LHO-4k & $(4.0 \pm 0.5)\times 10^{-22}$ & $(2.4 \pm 0.2)\times 10^{-22}$ \\ \hline
\end{tabular}
\end{center}
\caption{\label{t1} Summary of the 95\% upper limits delivered by
the two analysis methods when applied to the four interferometers
taking part in S1. GEO is the UK/German GEO\,600 interferometer in
Hanover, LLO is the LIGO instrument in Livingston, LA. LHO-2k and
LHO-4k are the two LIGO interferometers co-sited at Hanford, WA. }
\end{table}
In principle the Bayesian analysis allows the data from all four
interferometers to be combined to give an improved joint upper
limit, although the relative phase calibration of the instruments
was not sufficiently reliable to do this for S1.  It is however
interesting to note that an upper limit for $h_0$ can, within a
suitable model, be interpreted as an upper limit on the equatorial
ellipticity of the neutron star.  If we take the distance to the
pulsar to be 3.6\,kpc, its radius to be 10\,km and its mass to be
$1.4\,M_\odot$ a value of $h_0<1.4\times10^{-22}$ corresponds to
an ellipticity of less than $2.7\times10^{-4}$.  Such a large
upper limit to the ellipticity could be rejected a priori, as it
implies a much greater spindown rate (via gravitational
luminosity) than is seen in this pulsar. However as detector
sensitivities improve we can expect the above methodology to
deliver constraints on neutron star parameters not otherwise
achievable.

\section{Acknowledgements}
The authors gratefully acknowledge the support of the United
States National Science Foundation for the construction and
operation of the LIGO Laboratory and the Particle Physics and
Astronomy Research Council of the United Kingdom, the
Max-Planck-Society and the State of Niedersachsen/Germany for
support of the construction and operation of the GEO\,600
detector. The authors also gratefully acknowledge the support of
the research by these agencies and by the Australian Research
Council, the Natural Sciences and Engineering Research Council of
Canada, the Council of Scientific and Industrial Research of
India, the Department of Science and Technology of India, the
Spanish Ministerio de Ciencia y Tecnologia, the John Simon
Guggenheim Foundation, the David and Lucile Packard Foundation,
the Research Corporation, and the Alfred P.\ Sloan Foundation.

 \References
\item[] The LIGO Scientific Collaboration: Abbott B, et al.\ 2003a {\it gr-qc}/0308043
\item[] The LIGO Scientific Collaboration: Abbott B, et al.\ 2003b {\it gr-qc}/0308050
\item[] Hough J, et al.\ 1983 {\it Nature} 303, 216
\item[] Hereld M 1983 PhD dissertation, California Institute of
Technology
\item[] Jaranowski P, Kr\'olak A and Schutz BF 1998 {\it Phys.\ Rev.\
D} 58, 063001
\endrefs
\end{document}